\documentclass[12pt]{article}

\usepackage{amssymb}
\usepackage{cite}

\makeatletter
\renewcommand\thesubsection{\thesection.\@arabic\c@subsection}
\makeatother

\newcommand{\sect}[1]{\setcounter{equation}{0}\section{#1}}

\newcommand {\beq}{\begin{equation}}
\newcommand {\eeq}{\end{equation}}
\newcommand {\beqa}{\begin{eqnarray}}
\newcommand {\eeqa}{\end{eqnarray}}         
\newcommand {\beqs}{\begin{eqnarray*}}
\newcommand {\eeqs}{\end{eqnarray*}}
\newcommand {\bds}{\begin{displaymath}}
\newcommand {\eds}{\end{displaymath}}
\newcommand {\n}{\nonumber\\}

\newcommand {\bebb}{\begin{thebibliography}}
\newcommand {\eebb}{\end{thebibliography}}      
\newcommand {\bbit}{\bibitem}

\def\ra{\rangle}

\def\journal#1&#2(#3){\unskip, \sl #1\ \bf #2 \rm(19#3) }
\def\andjournal#1&#2(#3){\sl #1~\bf #2 \rm (19#3) }

\begin{document}

\newtheorem{Proposition}{Proposition}[section]
\newtheorem{Theorem}[Proposition]{Theorem}
\newtheorem{Definition}[Proposition]{Definition}
\newtheorem{Corollary}[Proposition]{Corollary}
\newtheorem{Lemma}[Proposition]{Lemma}
\newtheorem{Example}[Proposition]{Example}
\newtheorem{Remark}[Proposition]{Remark}


\begin{flushright}
\end{flushright}

\vskip 1cm

\begin{center}
{\Large\bf Hidden $sl(2)$-algebraic structure in Rabi model and its 2-photon and two-mode generalizations}

\vspace{1cm}

{\large Yao-Zhong Zhang}
\vskip.1in

{\em School of Mathematics and Physics, The University of Queensland,\\ Brisbane, Qld 4072, Australia}

{\em CAS Key Laboratory of Theoretical Physics, Institute of Theoretical Physics,\\ Chinese Academy of Sciences, Beijing 100190, China}

\end{center}

\date{}



\begin{abstract}
It is shown that the (driven) quantum Rabi model and its 2-photon and 2-mode generalizations possess a hidden
$sl(2)$-algebraic structure which explains the origin of the quasi-exact solvability of these models.
It manifests the first appearance of a hidden algebraic structure in quantum spin-boson systems without $U(1)$ symmetry.
\end{abstract}

\vskip.1in





\setcounter{section}{0}
\setcounter{equation}{0}
\sect{Introduction}

Quantum Rabi model and its multi-quantum and multi-mode generalizations constitute an important class of
spin-boson systems without $U(1)$ symmetry. They are phenomenological or theoretical systems used to model the ubiquitous matter-light
interactions in modern physics, and have applications in a variety of physical fields,
including quantum optics \cite{Vedral06}, cavity and circuit quantum electrodynamics \cite{Englund07,Niemczyk10}, solid state semiconductor systems \cite{Khitrova06} and trapped ions \cite{Leibfried03}.

The main difficulty in dealing with these models comes from the fact that not all their spectra seem algebraically accessible.
Majority parts of the spectra (i.e. the so-called regular energies) are given by the zeros of transcendental functions
which are either infinite power series or continued fractions with coefficients satisfying three-term recurrence relations \cite{Braak11,Moroz12,Chen12,Zhong13,Moroz13,Zhang13c,Moroz14,Zhong14,Zhang15,Duan15,Duan16}.
The exact locations of the zeros and thus closed-form expressions for the regular energies can not be determined via algebraic means.

It is well-known that under certain circumstances the Rabi model and its 2-photon and 2-mode generalizations admit exact, analytic solutions \cite{Reik82,Kus85,Emary02a,Emary02b,Zhang13a,Tomka14,Dossa14,Li15},
yielding closed-form expressions for parts of the energy spectra of the systems.
These ``exceptional" energies appear only when the model parameters satisfy some constraints. Thus the Rabi model and its 2-photon and 2-mode generalizations are quasi-exactly solvable \cite{Moroz13,Zhang13a}.

Quasi-exactly solvable systems are quantum mechanical problems for which only a finite part of their spectra can be found exactly
\cite{Turbiner88,Turbiner94,Ushveridze94,Gonzarez93}. They occupy an intermediate place between exactly solvable and non-solvable models. A typical feature of quasi-exact
solvability is the existence of a hidden algebraic structure.
The main purpose of this paper is to show that the Rabi model and its 2-photon and 2-mode generalizations possess a hidden $sl(2)$ structure,
i.e. they allow for an $sl(2)$ algebraization.
To our knowledge, this marks the first appearance of a hidden algebraic structure in quantum spin-boson models which do not have $U(1)$ symmetry.



\sect{General results}\label{exact-quasi-exact}

In this section we recall a general algebraic construction of quasi-exactly solvable differential equations \cite{Turbiner94}, and prove that the 2nd-order differential operator (\ref{Diff}) below has a hidden $sl(2)$ structure if its coefficients are algebraically dependent.

Let us take the $sl(2)$ algebra realized by the 1st-order differential operators in single variable $z$
\beq
J^+=z^2\frac{d}{dz}-nz,~~~~~J^0=z\frac{d}{dz}-\frac{n}{2},~~~~~J^-=\frac{d}{dz}.\label{Diff-JJJ}
\eeq
These differential operators satisfy the $sl(2)$ commutation relations for any value of the parameter $n$.
If $n$ is a non-negative integer, $n=0,1,2,\cdots,$ then (\ref{Diff-JJJ}) provide a
$(n+1)$-dimensional irreducible
representation ${\cal P}_{n+1}(z)={\rm span}\{1,z,z^2,\cdots,z^n\}$ of the $sl(2)$ algebra.
It is evident that any differential operator which is a polynomial of the $sl(2)$ generators (\ref{Diff-JJJ}) with $n$ being non-negative
integer will have the space ${\cal P}_{n+1}(z)$ as its invariant subspace, i.e. possesses $(n+1)$ eigen-functions in
the form of polynomial in $z$ of degree $n$. This is the main idea in \cite{Turbiner94} behind quasi-exact solvability of a differential operator.
Such differential operators are said to have a hidden $sl(2)$ algebraic structure
or allow for an $sl(2)$ algebraization.

Now consider the 2nd order differential operator of the form
\beq
{\cal H}=X(z)\frac{d^2}{dz^2}+Y(z)\frac{d}{dz}+Z(z),\label{Diff}
\eeq
where $X(z), Y(z), Z(z)$ are polynomials of degree at most 4, 3, 2 respectively,
$$
X(z)=\sum_{k=0}^4a_kz^k,~~~~~Y(z)=\sum_{k=0}^3b_kz^k,~~~~~Z(z)=\sum_{k=0}^2c_kz^k.
$$
The differential operator (\ref{Diff}) is usually called the Heun operator.
Then we have
\begin{Proposition}\label{General-Thm}
The differential operator ${\cal H}$ allows for an $sl(2)$ algebraization, i.e. has a hidden $sl(2)$ algebraic structure, if and only if
\beq
b_3=-2(n-1)a_4,~~~~c_2=n(n-1)a_4,~~~~c_1=-n[(n-1)a_3+b_2].\label{a3c2c1}
\eeq
\end{Proposition}
\noindent {\it Proof.} It suffices to prove that ${\cal H}$ is a quadratic combination of the $sl(2)$ generators (\ref{Diff-JJJ}) if and only if the relations (\ref{a3c2c1}) are satisfied.

\underline{Sufficiency}. We have
\beqa
{\cal H}&=&X(z)\frac{d^2}{dz^2}+\left[-2(n-1)a_4z^3+b_2z^2+b_1z+b_0\right]\frac{d}{dz}\n
& &  +n(n-1)a_4z^2-n[(n-1)a_3+b_2]z+c_0.\label{Suff-H1}
\eeqa
It is easy to check that
\beqa
&&a_4J^+J^++a_3J^+J^0+a_2J^0J^0+a_1J^0J^-+a_0J^-J^-\n
&&~~~~~~=X(z)\frac{d^2}{dz^2}+\left[-2(n-1)a_4z^3-\frac{3n-2}{2}a_3z^2-(n-1)a_2z-\frac{n}{2}a_1\right]\frac{d}{dz}\n
&&~~~~~~~~~~+n(n-1)a_4z^2+\frac{n^2}{2}a_3z+\frac{n^2}{2}a_2,\n
&&b_2J^++b_1J^0+b_0J^-=(b_2z^2+b_1z+b_0)\frac{d}{dz}-nb_2z-\frac{n}{2}b_1.
\eeqa
Substituting into (\ref{Suff-H1}) gives rise to
\beqa
{\cal H}&=&a_4J^+J^++a_3J^+J^0+a_2J^0J^0+a_1J^0J^-+a_0J^-J^-+\left(\frac{3n-2}{2}a_3+b_2\right)J^+\n
& &+[(n-1)a_2+b_1]J^0+\left(\frac{n}{2}a_1+b_0\right)J^-+\frac{n}{2}\left[\left(\frac{n}{2}-1\right)a_2+b_1\right]
   +c_0\label{Suff-H2}
\eeqa

\underline{Necessity}. We take
\beqa
{\cal H}&=&A_{++}J^+J^++A_{+0}J^+J^0+A_{00}J^0J^0+A_{0-}J^0J^-\n
& &+A_{--}J^-J^-+A_+J^++A_0J^0+A_-J^-+A_* ,\label{Nec-H1}
\eeqa
where $A_{++}$ etc are constant coefficients to be determined. Then by means of the expressions (\ref{Diff-JJJ}),
\beqa
{\cal H}&=&\left(A_{++}z^4+A_{+0}z^3+A_{00}z^2+A_{0-}z+A_{--}\right)\frac{d^2}{dz^2}
  + \left[-2(n-1)A_{++}z^3\right.\n
& &\left.+\left(A_+-\frac{3n-2}{2}A_{+0}\right)z^2
+\left(A_0-(n-1)A_{00}\right)z+A_--\frac{n}{2}\right]\frac{d}{dz}\n
& &+n(n-1)A_{++}z^2+n\left(\frac{n}{2}A_{+0}-A_+\right)z+\frac{n}{2}\left(\frac{n}{2}-A_0\right)+A_*.\label{Nec-H2}
\eeqa
The r.h.s. of (\ref{Nec-H2}) can be written as $X(z)\frac{d^2}{dz^2}+Y(z)\frac{d}{dz}+Z(z)$ provided that we make the identification
\beqa
&&a_4=A_{++},~~~~a_3=A_{+0},~~~~a_2=A_{00},~~~~a_1=A_{0-},~~~~a_0=A_{--},\n
&&b_3=-2(n-1)A_{++},~~~~b_2=A_+-\frac{3n-2}{2}A_{+0},~~~~b_1=A_0-(n-1)A_{00},\n
&&b_0=A_--\frac{n}{2},~~~~c_2=n(n-1)A_{++},~~~~c_1=n\left(\frac{n}{2}A_{+0}-A_+\right),\n
&&c_0=\frac{n}{2}\left(\frac{n}{2}-A_0\right)+A_*.
\eeqa
It follows that
\beq
b_3=-2(n-1)a_4,~~~~c_2=n(n-1)a_4,~~~~c_1=-n[(n-1)a_3+b_2].
\eeq
This completes our proof. \hfill $\Box$

\vskip.1in
In the following sections, we will apply the general results in Proposition \ref{General-Thm} to show that the (driven) Rabi model and its 2-photon and 2-mode generalizations possess a hidden $sl(2)$ algebraic structure.

\sect{Hidden $sl(2)$ structure in (driven) Rabi model}\label{rabi}

The Hamiltonian of the driven Rabi model is
\begin{equation}
H_R=\omega a^\dagger a+\Delta\sigma_z+g\,\sigma_x\left[a^\dagger+a\right]+\delta\,\sigma_x,\label{RabiH}
\end{equation}
where $g$ is the interaction strength,
$\sigma_z, \sigma_x$ are the Pauli matrices describing the two atomic levels separated
by energy difference $2\Delta$, and $a^\dagger$ ($a$) are
creation (annihilation) operators of a boson mode with frequency $\omega$. Here $a^\dagger$ ($a$) satisfy the Heisenberg algebra relations $[a, a^\dagger]=1,~~[a,a]=0=a^\dagger, a^\dagger]$.
The addition of the driving term $\delta \sigma_x$ breaks the $Z_2$ symmetry of the Rabi model. The driven Rabi model (\ref{RabiH}) is relevant to the description of some hybrid mechanical systems (see e.g. \cite{Zhong14}).

By means of the Fock-Bargmann correspondence $a^\dagger\rightarrow z,~a\rightarrow\frac{d}{dz}$,
the Hamiltonian becomes a matrix differential operator
\beq
H_R=\omega z\frac{d}{dz}+\Delta \sigma_z+g\,\sigma_x\left(z+\frac{d}{dz}\right)+\delta\,\sigma_x.
\eeq
Working in a representation defined by $\sigma_x$ diagonal and in terms of the two-component wavefunction
$\psi(z)=\left(\begin{array}{c}
\psi_+(z)\\
\psi_-(z)
\end{array} \right)$,
the time-independent Schr\"odinger equation $H_R\psi(z)=E\psi(z)$ gives rise to a coupled system of two 1st-order differential equations
\beqa
&&(\omega z+g)\frac{d}{dz}\psi_+(z)+[gz-(E-\delta)]\psi_+(z)+\Delta\psi_-(z)=0,\n
&&(\omega z-g)\frac{d}{dz}\psi_-(z)-[gz+(E+\delta)]\psi_-(z)+\Delta\psi_+(z)=0.\label{Rabi-matrix-diff}
\eeqa
If $\Delta=0$ these two equations decouple and reduce to the differential equations of two uncoupled displaced harmonic oscillators  \cite{Zhang13b}.
For this reason we will concentrate on the non-trivial $\Delta\neq 0$ case.

With the substitution $\psi_\pm(z)=e^{-gz/\omega}\phi_\pm(z)$, it follows
\beqa
&&\left[(\omega z+g)\frac{d}{dz}-\left(\frac{g^2}{\omega}-\delta+E\right)\right]\phi_+(z)=-\Delta\phi_-(z),\n
&&\left[(\omega z-g)\frac{d}{dz}-\left(2gz-\frac{g^2}{\omega}+\delta+E\right)\right]\phi_-(z)=-\Delta\phi_+(z).
  \label{Rabi-diff}
\eeqa
Eliminating $\phi_-(z)$ from the system we obtain the uncoupled differential equation for $\phi_+(z)$,
\beq
{\cal H}_R\phi_+(z)=\Delta^2\phi_+(z),\label{Spectral-eqn-Rabi}
\eeq
where
\beqa
{\cal H}_R&=&(\omega z-g)(\omega z+g)\frac{d^2}{dz^2}+\left[-2\omega gz^2+(\omega^2-2g^2-2E\omega)z-g\omega\right.\n
& &\left.   +2g\left(\frac{g^2}{\omega}-\delta\right)\right]\frac{d}{dz}
  +2g\left(\frac{g^2}{\omega}-\delta+E\right)z+E^2-\left(\delta-\frac{g^2}{\omega}\right)^2.\label{Rabi-H}
\eeqa

By Proposition (\ref{General-Thm}), ${\cal H}_R$ allows for an $sl(2)$ algebraization if
\beq
2g\left(E+\frac{g^2}{\omega}-\delta\right)\equiv c_1=-n[(n-1)a_3+b_2]\equiv 2g\omega n,
\eeq
which gives one set of the exact (exceptional) energies of the driven Rabi model
\beq
E=\omega n+\delta-\frac{g^2}{\omega},~~~~~n=0,1,2,\cdots.\label{Rabi-solution-E}
\eeq
Indeed, for such $E$ values, ${\cal H}_R$ is dependent on the integer parameter $n$ and can be expressed as the quadratic combination of the $sl(2)$ generators (\ref{Diff-JJJ})
\beqa
{\cal H}_R&=&\omega^2 J^0J^0-g^2 J^-J^--2g\omega J^++(n\omega^2-2g^2-2\omega E)J^0\n
& &-g\left[\omega+2\left(\delta-\frac{g^2}{\omega}\right)\right]J^-+n\left(\frac{n}{4}
\omega^2  -g-\omega E\right)+E^2-\left(\delta-\frac{g^2}{\omega}\right)^2,\label{Algebraization-Rabi}
\eeqa
where $E$ is given by ({\ref{Rabi-solution-E}).

Similarly for the other set of solutions of the driven Rabi model, we set
$\psi_\pm(z)=e^{gz/\omega}\varphi_\pm(z)$ and get from (\ref{Rabi-matrix-diff})
\beqa
&&\left[(\omega z+g)\frac{d}{dz}+\left(2gz+\frac{g^2}{\omega}+\delta-E\right)\right]\varphi_+(z)=-\Delta\varphi_-(z),\n
&&\left[(\omega z-g)\frac{d}{dz}-\left(\frac{g^2}{\omega}+\delta+E\right)\right]\varphi_-(z)=-\Delta\varphi_+(z).
  \label{Rabi-diff'}
\eeqa
Eliminating $\varphi_+(z)$ from the system we obtain the uncoupled differential equation for $\varphi_-(z)$,
\beq
\tilde{\cal H}_R\varphi_-(z)=\Delta^2\varphi_-(z),\label{Spectral-eqn-Rabi'}
\eeq
where
\beqa
\tilde{\cal H}_R&=&(\omega z-g)(\omega z+g)\frac{d^2}{dz^2}+\left[2\omega gz^2+(\omega^2-2g^2-2E\omega)z+g\omega\right.\n
& &\left.   -2g\left(\frac{g^2}{\omega}+\delta\right)\right]\frac{d}{dz}
  +2g\left(\frac{g^2}{\omega}+\delta+E\right)z+E^2-\left(\delta+\frac{g^2}{\omega}\right)^2.\label{Rabi-H'}
\eeqa

$\tilde{\cal H}_R$ allows for an $sl(2)$ algebraization if
\beq
-2g\left(E+\frac{g^2}{\omega}+\delta\right)\equiv c_1=-n[(n-1)a_3+b_2]\equiv -2g\omega n,
\eeq
which gives the other set of the exact (exceptional) energies of the driven Rabi model
\beq
E=\omega n-\delta-\frac{g^2}{\omega},~~~~~n=0,1,2,\cdots.\label{Rabi-solution-E'}
\eeq
For such $E$ values, $\tilde{\cal H}_R$ is dependent on the integer parameter $n$ and can be expressed as the quadratic combination of the $sl(2)$ generators (\ref{Diff-JJJ})
\beqa
\tilde{\cal H}_R&=&\omega^2 J^0J^0-g^2 J^-J^-+2g\omega J^++(n\omega^2-2g^2-2\omega E)J^0\n
& &+g\left[\omega-2\left(\delta+\frac{g^2}{\omega}\right)\right]J^-+n\left(\frac{n}{4}
\omega^2  -g-\omega E\right)+E^2-\left(\delta+\frac{g^2}{\omega}\right)^2,\label{Algebraization-Rabi'}
\eeqa
where $E$ is given by ({\ref{Rabi-solution-E'}).

Some remarks are in order. Exceptional energies (\ref{Rabi-solution-E}) and ({\ref{Rabi-solution-E'}) coincide with
those obtained by other approaches (see e.g. the appendix of \cite{Zhang13c}, and \cite{Zhong14,Li15}). The $sl(2)$ algebraizations (\ref{Algebraization-Rabi}) and (\ref{Algebraization-Rabi'}) mean that the corresponding spectral problems (\ref{Spectral-eqn-Rabi}) and (\ref{Spectral-eqn-Rabi'})
possess $(n+1)$ eigenfunctionns, respectively, in the form of polynomials of degree $n$.  Other eigenfunctions are non-polynomial and
are in general given by infinite power series with coefficients satisfying three-term recurrence relations \cite{Braak11,Moroz12,Zhong13,Zhang13c,Zhong14}.

\sect{Hidden $sl(2)$ algebraic structure in 2-photon Rabi model}\label{2-photon}
The Hamiltonian of the 2-photon Rabi model reads
\begin{equation}
H_{2-p}=\omega a^\dagger a+\Delta\sigma_z+g\,\sigma_x\left[(a^\dagger)^2+a^2\right],\label{2-photon-RabiH1}
\end{equation}
where $g$ is the interaction strength.
Introduce the operators $K_\pm, K_0$ by
\beq
K_+=\frac{1}{2}(a^\dagger)^2,~~~~K_-=\frac{1}{2}a^2,~~~~K_0=\frac{1}{2}
\left(a^\dagger a+\frac{1}{2}\right).   \label{2-photon-Rabi-boson}
\eeq
Then the Hamiltonian (\ref{2-photon-RabiH1}) becomes
\beq
H_{2-p}=2\omega\left(K_0-\frac{1}{4}\right)+\Delta\sigma_z+2g\,\sigma_x(K_++K_-).\label{2-photon-RabiH2}
\eeq
The operators $K_\pm, K_0$ form the usual $su(1,1)$ Lie algebra,
\beq
[K_0, K_\pm]=\pm K_\pm,~~~~~ [K_+, K_-]=-2K_0.\label{su11-relations}
\eeq
The quadratic Casimir operator $C$ of the algebra is given by
\beq
C=K_+K_--K_0(K_0-1).\label{su11-casimir}
\eeq

Consider the infinite-dimensional unitary irreducible representation of
$su(1,1)$ known as the positive discrete series ${\cal D}^+(q)$, where
the parameter $q$ is the so-called Bargmann index. In this representation the basis states $\{|q,n\ra\}$ diagonalize the operator $K_0$,
\beq
K_0|q,n\ra=(n+q)|q,n\ra\label{K0-rep}
\eeq
for $q>0$ and $n=0,1,2, \cdots$, and the Casimir operator $C$ has the eigenvalue $q(1-q)$.
The operators $K_+$ and $K_-$ are hermitian to each other
and act as raising and lowering operators respectively within ${\cal D}^+(q)$,
\beqa
K_+|q,n\ra&=&\sqrt{(n+1)(n+2q)}\;|q,n+1\ra,\n
K_-|q,n\ra&=&\sqrt{n(n+2q-1)}\;|q,n-1\ra.\label{su11-rep}
\eeqa
It is well-known that the single-mode bosonic realization (\ref{2-photon-Rabi-boson}) provides a representation of ${\cal D}^+(q)$ with $C=\frac{3}{16}$ and
$q=\frac{1}{4}, \frac{3}{4}$.

By means of the Fock-Bargmann correspondence the operators $K_\pm, K_0$ (\ref{2-photon-Rabi-boson}) are realized by
single-variable 2nd-order differential operators
\beq
{K}_0 = z\frac{d}{dz}+ q, ~~~~{K}_+ = \frac{z}{2}, ~~~~
{K}_- =2z\frac{d^2}{dz^2}+4q\frac{d}{dz},\label{2-photon-Rabi-diff}
\eeq
and the 2-photon Rabi Hamiltonian becomes \cite{Zhang13a}
\beq
H_{2-p}=2\omega\left(z\frac{d}{dz}+q-\frac{1}{4}\right)+\Delta\sigma_z+2g\,\sigma_x
  \left(\frac{z}{2}+2z\frac{d^2}{dz^2}+4q\frac{d}{dz}\right).
\eeq
Working in a representation defined by $\sigma_x$ diagonal and in terms of the two component wavefunction,
$\psi(z)=\left(\begin{array}{c}
\psi_+(z)\\
\psi_-(z)
\end{array} \right)$,
the time-independent Schr\"odinger equation $H_{2-p}\psi(z)=E\psi(z)$ leads to two coupled 2nd-order differential equations,
\beqa
&&4gz\frac{d^2}{dz^2}\psi_+(z)+(2\omega z+8gq)\frac{d}{dz}\psi_+(z)
   +\left[gz+2\omega\left(q-\frac{1}{4}\right)-E\right]\psi_+(z)+\Delta\psi_-(z)=0,\n
&&4gz\frac{d^2}{dz^2}\psi_-(z)+(-2\omega z+8gq)\frac{d}{dz}\psi_-(z)
   +\left[gz-2\omega\left(q-\frac{1}{4}\right)+E\right]\psi_-(z)-\Delta\psi_+(z)=0.\n
\eeqa
If $\Delta=0$ these equations reduce to the differential equations of two uncoupled single-mode squeezed harmonic oscillators \cite{Zhang13b}.
In the following we will concentrate on the $\Delta\neq 0$ case.

With the substitution
\beq
\psi_\pm(z)=e^{-\frac{\omega}{4g}  (1-\Omega) z}\varphi_\pm(z),~~~~~~\Omega=\sqrt{1-\frac{4g^2}{\omega^2}},\label{2-photon-sustitution}
\eeq
where $\left|\frac{2g}{\omega}\right|<1$, it follows \cite{Zhang13a}
\beqa
&&\left\{4gz\frac{d^2}{dz^2}+[2\omega\Omega z+8gq]\frac{d}{dz}
  +2q\omega\Omega-\frac{1}{2}\omega-E\right\}\varphi_+=-\Delta \varphi_-,\n
&&\left\{4gz\frac{d^2}{dz^2}+[2\omega(\Omega-2)z+8gq]\frac{d}{dz}
  +\frac{\omega^2}{g}(1-\Omega)z+ 2q\omega(\Omega-2)+\frac{1}{2}\omega+E\right\}\varphi_-=\Delta \varphi_+.\n
  \label{2-photon-diff}
\eeqa
Eliminating $\varphi_-(z)$ from the system, we obtain the 4th-order differential equation for $\varphi_+(z)$
\beq
{\cal H}_{2-p}\varphi_+(z)=-\Delta^2\varphi_+(z),\label{Spectral-eqn-2photon}
\eeq
where
\beqa
{\cal H}_{2-p}&=&16g^2z^2\frac{d^4}{dz^4}+64g^2\left[\frac{\omega}{4g}(\Omega-1) z^2+\left(q+\frac{1}{2}\right)z\right]\frac{d^3}{dz^3}\n
& &+\left\{4\omega^2(\Omega^2-3\Omega+1)z^2+16\omega g\left[3\left(q+\frac{1}{2}\right)\Omega-3q-1\right]z+64g^2q\left(q+\frac{1}{2}\right)\right\}\frac{d^2}{dz^2}\n
& &+\left\{2\frac{\omega^3}{g}\Omega(1-\Omega)z^2+\left[8\omega^2q(1-\Omega)+8\omega^2\left(q+\frac{1}{2}\right)(1-\Omega)^2\right.\right.\n
& &~~~~~\left.\left.     +4\omega\left(E-2\omega\left(q+\frac{1}{4}\right)\right)\right]z  +32\omega gq\left[\left(q+\frac{1}{2}\right)\Omega-q\right]\right\}\frac{d}{dz}\n
& &+\frac{\omega^2}{g}(1-\Omega)\left(2q\omega\Omega-\frac{1}{2}\omega-E\right)z
 +4\omega^2q^2(1-\Omega)^2-\left[E-2\omega\left(q-\frac{1}{4}\right)\right]^2.\label{2-photon-H1}
\eeqa
Using the identities
\beq
z^2\frac{d^4}{dz^4}=J^+(J^-)^3+nz\frac{d^3}{dz^3},~~~~
z^2\frac{d^3}{dz^3}=J^+(J^-)^2+nz\frac{d^2}{dz^2},~~~~
z\frac{d^3}{dz^3}=J^0(J^-)^2+\frac{n}{2}\frac{d^2}{dz^2},\label{Diff-identities}
\eeq
we obtain
\beq
{\cal H}_{2-p}=16g^2J^+(J^-)^3+16g\omega(\Omega-1)J^+(J^-)^2
  +16g^2[n+2(2q+1)]J^0(J^-)^2+{\cal H}^{(2)}_{2-p},
\eeq
where
\beqa
{\cal H}^{(2)}_{2-p}&=&\left\{4\omega^2(\Omega^2-3\Omega+1)z^2
+16\omega g\left[(\Omega-1)n+3\left(q+\frac{1}{2}\right)\Omega-3q-1\right]z\right.\n
& &\left.+8g^2n\left[n+4\left(q+\frac{1}{2}\right)\right]
  +64g^2q\left(q+\frac{1}{2}\right)\right\}\frac{d^2}{dz^2}\n
& &+\left\{2\frac{\omega^3}{g}\Omega(1-\Omega)z^2+\left[8\omega^2q(1-\Omega)+8\omega^2\left(q+\frac{1}{2}\right)(1-\Omega)^2\right.\right.\n
& &~~~~~\left.\left.     +4\omega\left(E-2\omega\left(q+\frac{1}{4}\right)\right)\right]z  +32\omega gq\left[\left(q+\frac{1}{2}\right)\Omega-q\right]\right\}\frac{d}{dz}\n
& &+\frac{\omega^2}{g}(1-\Omega)\left(2q\omega\Omega-\frac{1}{2}\omega-E\right)z
 +4\omega^2q^2(1-\Omega)^2-\left[E-2\omega\left(q-\frac{1}{4}\right)\right]^2.\n
 \label{2-photon-H2}
\eeqa

${\cal H}^{(2)}_{2-p}$ allows for an $sl(2)$ algebraization if
\beqa
\frac{\omega^2}{g}(1-\Omega)\left(2q\omega\Omega-\frac{1}{2}\omega-E\right)&\equiv&
   c_1=-n[(n-1)a_3+b_2]\n
&\equiv&-2\frac{\omega^2}{g}(1-\Omega)\Omega n,
\eeqa
which, for $\Omega\neq 1$ (the $\Omega=1$ case is trivial as it corresponds to $g=0$), gives the exact (exceptional) energies of the 2-photon Rabi model
\beqa
&&E=-\frac{1}{2}\omega+\left[2n+2\left(q-\frac{1}{4}\right)+\frac{1}{2}\right]\omega\Omega,~~~~~n=0,1,2,\cdots.\label{2-photon-energy}
\eeqa
Indeed for such $E$ values ${\cal H}_{2-p}$ depends on the integer parameter $n$ and can be expressed as the quartic combination of the $sl(2)$ generators  (\ref{Diff-JJJ})
\beqa
{\cal H}_{2-p}&=&16g^2J^+(J^-)^3+16g\omega(\Omega-1)J^+(J^-)^2
  +16g^2[n+2(2q+1)]J^0(J^-)^2\n
& &+4\omega^2(\Omega^2-3\Omega+1)J^0J^0
  +16\omega g\left[(\Omega-1)n+3\left(q+\frac{1}{2}\right)\Omega-3q-1\right]J^0J^-\n
& &\left[+8g^2n\left(n+4(q+\frac{1}{2})\right)
  +64g^2q\left(q+\frac{1}{2}\right)\right]J^-J^-
  +2\frac{\omega^3}{g}\Omega(1-\Omega)J^+\n
& &+\left[4\omega^2(n-1)(\Omega^2-3\Omega+1)+8\omega^2q(1-\Omega)\right.\n
& &~~~\left.+8\omega^2\left(q+\frac{1}{2}\right)(1-\Omega)^2
  +4\omega\left(E-2\omega(q+\frac{1}{4})\right)\right]J^0\n
& &\left[+8g\omega n\left((\Omega-1)n+3(q+\frac{1}{2})\Omega-3q-1\right)
   +32\omega g q\left((q+\frac{1}{2})\Omega-q\right)\right]J^-\n
& &+n(n-2)\omega^2(\Omega^2-3\Omega+1)+4n\omega^2 q(1-\Omega)
   +4n\omega^2\left(q+\frac{1}{2}\right)(1-\Omega)^2\n
& &+\omega n\left[E-2\omega\left(q+\frac{1}{4}\right)\right]
  +4\omega^2 q^2(1-\Omega)^2-\left[E-2\omega\left(q-\frac{1}{4}\right)\right]^2.
\eeqa
Here $E$ is given by (\ref{2-photon-energy}). This $sl(2)$ algebraization demonstrates that for each energy value $E$ in (\ref{2-photon-energy}) the 2-photon Rabi model has a hidden $sl(2)$ algebraic structure.

Notice that the exceptional energies (\ref{2-photon-energy}) coincide with those obtained in \cite{Emary02a,Zhang13a} via different methods.
It is clear that the corresponding spectral problem (\ref{Spectral-eqn-2photon}) has $n+1$ polynomial eigenfunctions in $z$ of degree $n$. Other eigenfunctions are non-polynomial and can not be obtained in closed analytic form.
We remark that as shown in \cite{Zhang15} when $\Omega=0$, i.e. $|2g/\omega|=1$, the 2-photon Rabi model has no entire solutions.

\sect{Hidden $sl(2)$ algebraic structure in two-mode Rabi model}\label{2-mode}

The Hamiltonian of the two-mode quantum Rabi model reads \cite{Zhang13a}
\beq
H_{2-m}=\omega(a_1^\dagger a_1+a_2^\dagger a_2)+\Delta\sigma_z+g\,\sigma_x(a_1^\dagger a_2^\dagger+a_1 a_2),
  \label{2-mode-RabiH1}
\eeq
where we assume that the boson modes are degenerate with the same frequency $\omega$ and $g$ is the coupling constant.
Introduce the operators $K_\pm, K_0$,
\beq
K_+=a_1^\dagger a_2^\dagger,~~~~K_-=a_1 a_2,~~~~K_0=\frac{1}{2}(a_1^\dagger a_1+a_2^\dagger a_2+1).\label{2-mode-Rabi-boson}
\eeq
Then the Hamiltonian (\ref{2-mode-RabiH1}) becomes \cite{Zhang13a}
\beq
H_{2-m}=2\omega\left(K_0-\frac{1}{2}\right)+\Delta\sigma_z+g\sigma_x(K_++K_-).\label{2-mode-RabiH2}
\eeq
The operators $K_\pm, K_0$ form the $su(1,1)$ algebra (\ref{su11-relations}).
As in the previous section we shall use the unitary irreducible representation (i.e. the positive discrete series). However, to avoid confusion in this section we shall use $\kappa$ to denote the Bargmann
index of the representation. Using this notation the action of the operators $K_\pm, K_0$ and the Casimir $C$ (\ref{su11-casimir}) on the basis states  $|\kappa,n\ra$ of the representation reads
\beqa
K_0|\kappa,n\ra&=&(n+\kappa)\;|\kappa,n\ra,\n
K_+|\kappa,n\ra&=&\sqrt{(n+2\kappa)(n+1)}\;|\kappa,n+1\ra,\n
K_-|\kappa,n\ra&=&\sqrt{(n+2\kappa-1)n}\;|\kappa,n-1\ra,\n
C|\kappa,n\ra&=&\kappa(1-\kappa)\;|\kappa,n\ra,\label{2-mode-rep}
\eeqa
for $\kappa>0$ and $n=0,1,2,\cdots$.
For the two-mode bosonic realization (\ref{2-mode-Rabi-boson}) of $su(1,1)$ that we require here the Casimir $C$ takes the value $C=\kappa(1-\kappa)$  with
the Bargmann index $\kappa$ being any positive integers or half-integers,
i.e. $\kappa=1/2, 1, 3/2,\cdots$.

Using the Fock-Bargmann correspondence the operators $K_\pm, K_0$
(\ref{2-mode-Rabi-boson}) have the single-variable differential realization \cite{Zhang13b},
\beq
K_0=z\frac{d}{dz}+\kappa,~~~~K_+=z,~~~~K_-=z\frac{d^2}{dz^2}+2\kappa\frac{d}{dz}
   \label{su11-diff-rep-2mode}
\eeq
and the Hamiltonian (\ref{2-mode-RabiH2}) can be expressed as the matrix differential operator \cite{Zhang13a}
\beq
H_{2-m}=2\omega\left(z\frac{d}{dz}+\kappa-\frac{1}{2}\right)+\Delta\sigma_z+g\,\sigma_x\left(z+z\frac{d^2}{dz^2}
   +2\kappa\frac{d}{dz}\right).   \label{2-mode-Rabi-diff}
\eeq
Working in a representation defined by $\sigma_x$ diagonal and in terms of the two-component
wavefunction $\psi(z)=\left(\begin{array}{c}
\psi_+(z)\\
\psi_-(z)
\end{array} \right)$, we see that the time-independent Schr\"odinger equation
$H_{2-m}\psi(z)=E\psi(z)$ yields the two coupled differential equations,
\beqa
&&gz\frac{d^2}{dz^2}\psi_+(z)+2(\omega z+g\kappa)\frac{d}{dz}\psi_+(z)
   +\left[gz+2\omega\left(\kappa-\frac{1}{2}\right)-E\right]\psi_+(z)+\Delta\psi_-(z)=0,\n
&&gz\frac{d^2}{dz^2}\psi_-(z)+2(-\omega z+g\kappa)\frac{d}{dz}\psi_-(z)
   +\left[gz-2\omega\left(\kappa-\frac{1}{2}\right)+E\right]\psi_-(z)-\Delta\psi_+(z)=0.\n
\eeqa
If $\Delta=0$ these reduce to the differential equations of two uncoupled
two-mode squeezed harmonic oscillators  \cite{Zhang13a}.
So in the following we will concentrate on the $\Delta\neq 0$ case.

With the substitution
\beq
\psi_\pm(z)=e^{-\frac{\omega}{g}  (1-\Lambda) z}\varphi_\pm(z),~~~~~~\Lambda=\sqrt{1-\frac{g^2}{\omega^2}},
\eeq
where $\left|\frac{g}{\omega}\right|<1$, it follows \cite{Zhang13a}
\beqa
&&\left\{gz\frac{d^2}{dz^2}+2[\omega\Lambda z+g\kappa]\frac{d}{dz}
  +2\kappa\omega\Lambda-\omega-E\right\}\varphi_+=-\Delta \varphi_-,\n
&&\left\{gz\frac{d^2}{dz^2}+2[\omega(\Lambda-2)z+g\kappa]\frac{d}{dz}
  +\frac{4\omega^2}{g}(1-\Lambda)z+ 2\kappa\omega(\Lambda-2)+\omega+E\right\}\varphi_-=\Delta \varphi_+.\n
  \label{2-mode-diff}
\eeqa
Eliminating $\varphi_-(z)$ from the system, we obtain the 4th-order differential equation for $\varphi_+(z)$,
\beq
{\cal H}_{2-m}\varphi_+(z)=-\Delta^2\varphi_+(z),\label{Spectral-eqn-2mode}
\eeq
where
\beqa
{\cal H}_{2-m}&=&g^2z^2\frac{d^4}{dz^4}+4g^2\left[\frac{\omega}{g}(\Lambda-1) z^2
   +\left(\kappa+\frac{1}{2}\right)z\right]\frac{d^3}{dz^3}\n
& &+\left\{4\omega^2(\Lambda^2-3\Lambda+1)z^2+4\omega g\left[3\left(\kappa+\frac{1}{2}\right)\Lambda-3\kappa-1\right]z
   +4g^2\kappa\left(\kappa+\frac{1}{2}\right)\right\}\frac{d^2}{dz^2}\n
&&+\left\{8\frac{\omega^3}{g}\Lambda(1-\Lambda)z^2+\left[8\omega^2\kappa(1-\Lambda)
   +8\omega^2\left(\kappa+\frac{1}{2}\right)(1-\Lambda)^2\right.\right.\n
&&~~~~~~\left.\left.     +4\omega(E-2\omega\kappa)\right]z
   +8\omega g\kappa\left[\left(\kappa+\frac{1}{2}\right)\Lambda-\kappa\right]\right\}\frac{d}{dz}\n
&&+4\frac{\omega^2}{g}(1-\Lambda)\left(2\kappa\omega\Lambda-\omega-E\right)z
+4\omega^2\kappa^2(1-\Lambda)^2-\left[E-2\omega\left(\kappa-\frac{1}{2}\right)\right]^2.\label{2-mode-H1}
\eeqa
By means of the identities (\ref{Diff-identities}) we have
\beq
{\cal H}_{2-m}=gJ^+(J^-)^3+4\omega g(\Lambda-1)J^+(J^-)^2
   +g^2\left[n+4(\kappa+\frac{1}{2})\right]J^-(J^-)^2+{\cal H}^{(2)}_{2-m},
\eeq
where
\beqa
{\cal H}^{(2)}_{2-m}&=&\left\{4\omega^2(\Lambda^2-3\Lambda+1)z^2+4\omega g\left[(\Lambda-1)n+3\left(\kappa+\frac{1}{2}\right)\Lambda-3\kappa-1\right]z\right.\n
& &\left.+g^2\frac{n}{2}\left[n+4(\kappa+\frac{1}{2})\right]
   +4g^2\kappa\left(\kappa+\frac{1}{2}\right)\right\}\frac{d^2}{dz^2}\n
&&+\left\{8\frac{\omega^3}{g}\Lambda(1-\Lambda)z^2+\left[8\omega^2\kappa(1-\Lambda)
   +8\omega^2\left(\kappa+\frac{1}{2}\right)(1-\Lambda)^2\right.\right.\n
&&~~~~~~\left.\left.     +4\omega(E-2\omega\kappa)\right]z
   +8\omega g\kappa\left[\left(\kappa+\frac{1}{2}\right)\Lambda-\kappa\right]
   \right\}\frac{d}{dz}\n
&&+4\frac{\omega^2}{g}(1-\Lambda)\left(2\kappa\omega\Lambda-\omega-E\right)z
+4\omega^2\kappa^2(1-\Lambda)^2-\left[E-2\omega\left(\kappa-\frac{1}{2}\right)\right]^2.\n
\label{2-mode-H2}
\eeqa

Similar to the 2-photon Rabi case, ${\cal H}^{(2)}_{2-m}$ allows for an $sl(2)$ algebraization if
\beqa
4\frac{\omega^2}{g}(1-\Lambda)\left(2\kappa\omega\Lambda-\omega-E\right)&\equiv&c_1
  =-n[(n-1)a_3+b_2]\n
&\equiv& -8\frac{\omega^3}{g}\Lambda(1-\Lambda)n,
\eeqa
which, for $\Lambda\neq 1$ (the $\Lambda=1$ case is trivial as it corresponds to $g=0$), give the exceptional energies of the 2-mode Rabi model
\beqa
&&E=-\omega+\left[2n+2\left(\kappa-\frac{1}{2}\right)+1\right]\omega\Lambda.
  \label{2-mode-energy}
\eeqa
For such $E$ values, ${\cal H}_{2-m}$ depends on integer parameter $n$ and possesses an algebraization in terms of the $sl(2)$ generators (\ref{Diff-JJJ})
\beqa
{\cal H}_{2-m}&=&gJ^+(J^-)^3+4\omega g(\Lambda-1)J^+(J^-)^2
   +g^2\left[n+4\left(\kappa+\frac{1}{2}\right)\right]J^-(J^-)^2\n
& &+4\omega^2(\Lambda^2-3\Lambda+1)J^0J^0+4\omega g\left[(\Lambda-1)n
  +3\left(\kappa+\frac{1}{2}\right)\Lambda-3\kappa-1\right]J^0J^-\n
& &+\left\{g^2\frac{n}{2}\left[n+4\left(\kappa+\frac{1}{2}\right)\right]
   +4g^2\kappa\left(\kappa+\frac{1}{2}\right)\right\}J^-J^-
   +8\frac{\omega^3}{g}\Lambda(1-\Lambda)J^+\n
& &+\left[4\omega^2(n-1)(\Lambda^2-3\Lambda+1)+8\omega^2\kappa(1-\Lambda)\right.\n
& &~~~\left.   +8\omega^2\left(\kappa+\frac{1}{2}\right)(1-\Lambda)^2
   +4\omega(E-2\omega\kappa)\right]J^0\n
& &+\left\{2n\omega g\left[(\Lambda-1)n+3\left(\kappa+\frac{1}{2}\right)  \Lambda-3\kappa-1\right]+8\omega g\kappa\left[\left(\kappa+\frac{1}{2}\right)
  \Lambda-\kappa\right] \right\}J^-\n
& &+n\left(n-2\right)\omega^2(\Lambda^2-3\Lambda+1)
  +4n\omega^2\kappa(1-\Lambda)+4n\omega^2\left(\kappa+\frac{1}{2}\right)(1-\Lambda)^2\n
& &+2n\omega(E-2\omega\kappa)+4\omega^2\kappa^2(1-\Lambda)^2
  -\left[E-2\omega\left(\kappa-\frac{1}{2}\right)\right]^2.\label{Algebraization-2mode}
\eeqa
Here $E$ is  given by (\ref{2-mode-energy}). Thus for each energy value $E$ in (\ref{2-mode-energy}) the 2-mode Rabi model has a hidden $sl(2)$ algebraic structure.

We remark that the exceptional energies (\ref{2-mode-energy}) coincide with those presented in \cite{Zhang13a} via the Bethe ansatz method \cite{Zhang12}. The $sl(2)$ algebraization of (\ref{Algebraization-2mode}) implies that the corresponding
spectral problem (\ref{Spectral-eqn-2mode}) possesses $n+1$ polynomial eigenfunctions of degree $n$. Other eigenfunctions are non-polynomial and can not be found in closed analytic form.
Note that as shown in \cite{Zhang15} when $\Lambda=0$, i.e. $|g/\omega|=1$, the two-mode Rabi model has no entire solutions.

\vskip.4in
\noindent {\bf\large Acknowledgments:}
I would like to thank Alexander Turbiner for encouraging me to publish the results in this paper.
This work was partially supported by the Australian Research Council through
Discovery Projects grant DP140101492.

\bebb{99}

\bbit{Vedral06}
V. Vedral, Modern foundations of quantum optics, Imperial College Press, London, 2006.

\bbit{Englund07}
D. Englund et al, Nature {\bf 450}, 857 (2007).

\bbit{Niemczyk10}
T. Niemczyk et al, Nature Phys. {\bf 6}, 772 (2010).

\bbit{Khitrova06}
G. Khitrova, H.M. Gibbs, M. Kira, S.W. Koch and A. Scherer, Nature Phys. {\bf 2}, 81 (2006).

\bbit{Leibfried03}
D. Leibfried, R. Blatt, C. Monroe and D. Wineland, Rev. Mod. Phys. {\bf 75}, 281 (2003).

\bbit{Braak11}
D. Braak, Phys. Rev. Lett. {\bf 107}, 100401 (2011).

\bbit{Moroz12}
A. Moroz, Europhys. Lett. {\bf 100}, 60010 (2012).

\bbit{Chen12} Q.H. Chen, C. Wang, S. He, T. Liu and K.L. Wang, Phys. Rev. A {\bf 86}, 023822 (2012).

\bbit{Zhong13}
H. Zhong, Q. Xie, M.T. Batchelor and C. Lee, J. Phys. A: Math. Theor. {\bf 46}, 415302 (2013).

\bbit{Moroz13}
A. Moroz, Ann. Phys. {\bf 338}, 319 (2013).

\bbit{Zhang13c}
Y.-Z. Zhang, Analytic solutions of 2-photon and two-mode Rabi models, arVix:1304.7827v2 [quant-ph].

\bbit{Moroz14}
A. Moroz,  Ann. Phys. {\bf 340}, 252 (2014).

\bbit{Zhong14}
H. Zhong, Q. Xie, X.-W. Guan, M.T. Batchelor, K. Gao and C. Lee,
J. Phys. A: Math. Theor. {\bf 47}, 045301 (2014).

\bbit{Zhang15}
Y.-Z. Zhang, On the two-mode and $k$-photon quantum Rabi models, arXiv:1507.03863v2 [quant-ph].

\bbit{Duan15}
L. Duan, S. He, D. Braak and Q.-H. Chen, Europhys. Lett. {\bf 112}, 34003 (2015).

\bbit{Duan16}
L. Duan, Y.-F. Xie, D. Braak and Q.-H. Chen, Two-photon Rabi model: Analytic solutions and spectral collapse, arXiv:1603.04503v1 [quant-ph].

\bbit{Reik82}
H.G. Reik, H. Nusser and L.A. Ribeiro, J. Phys. A: Math. Gen. {\bf 15}, 3431 (1982).

\bbit{Kus85}
M. Kus, J. Math. Phys. {\bf 26}, 2792 (1985).


\bbit{Emary02a}
C. Emary and R.F. Bishop, J. Phys. A: Math. Gen. {\bf 35}, 8231 (2002).

\bbit{Emary02b}
C. Emary and R.F. Bishop, J. Math. Phys. {\bf 43}, 3916 (2002).

\bbit{Zhang13a}
Y.-Z. Zhang, J. Math. Phys. {\bf 54}, 102104  (2013).

\bbit{Tomka14}
 M. Tomka, O. El Araby, M. Pletyukhov and V. Gritsev, Phys. Rev. A {\bf 90}, 063839 (2014).

\bbit{Dossa14}
A.F. Dossa and G.Y.H. Avossevou, J. Math. Phys. {\bf 55}, 102104 (2014).

\bbit{Li15}
Z.-M. Li and  M.T. Batchelor,  J. Phys. A: Math. Theor. {\bf 48}, 454005 (2015).


\bbit{Turbiner88}
A. Turbiner, Comm. Math. Phys. {\bf 118}, 467 (1988).

\bbit{Turbiner94}
A. Turbiner, Quasi-exactly-solvable differential equations, in
CRC Handbook of Lie Group Analysis of Differential Equations,
Vol. {\bf 3}: New Trends in Theoretical Developments and Computational Methods,
Chapter 12, CRC Press, N. Ibragimov (ed.), pp. 331-366 (1995).



\bbit{Ushveridze94}
A.G. Ushveridze, Quasi-exactly solvable models in quantum mechanics, Institute of
Physics Publishing, Bristol, 1994.

\bbit{Gonzarez93}
A. Gonz\'arez-L\'opez, N. Kamran and P. Olver, Comm. Math. Phys. {\bf 153}, 117 (1993).

\bbit{Zhang13b}
Y.-Z. Zhang, J. Phys. A: Math. Theor. {\bf 46}, 455302 (2013).

\bbit{Zhang12}
Y.-Z. Zhang, J. Phys. A: Math. Theor. {\bf 45}, 065206 (2012).


\eebb

\end{document}